\documentclass{elsart}
\usepackage{graphicx}
\usepackage{amssymb}
\journal{New Astronomy}
\begin{document}
\begin{frontmatter}

 \title{A Far Infrared Polarimeter}
 \author{\normalsize{A.Catalano, L.Conversi, S.De Gregori, M.De Petris, L.Lamagna}}
 \author{\normalsize{R.Maoli, G.Savini}}
 \address{Dipartimento di Fisica, Universit\`{a} di Roma "La Sapienza", P.le Aldo Moro 2, Roma, Italia}

 \author{\normalsize{E.S.Battistelli}}
 \address{Instituto de Astrofisica de Canarias, Via Lactea, La Laguna Tenerife, Espa\~{n}a}

 \author{\normalsize{A.Orlando}}
 \address{Department of Physics and Astronomy, University of Wales, Cardiff}
 \thanks[author]{Corresponding authors: A.Catalano and S.De Gregori\\ (E-mail:~andrea.catalano@roma1.infn.it, simone.degregori@roma1.infn.it)}

 \begin{abstract}

 We describe an experiment to measure calibration sources, the polarization of Cosmic
 Microwave Background Radiation (CMBR) and the polarization induced on the CMBR from S-Z
 effects,
 using a polarimeter, MITOPol, that will be employed at the MITO
 telescope.
 \\
 Two modulation methods are presented and compared: an amplitude
 modulation with a Fresnel double rhomb and a phase modulation
 with a modified Martin-Puplett interferometer. A first light is
 presented from the campaign (summer 2003) that has permitted to
 estimate the instrument spurious polarization using the second modulation method.
 \end{abstract}
 \end{frontmatter}

\section{Introduction}

The Cosmic Microwave Background Polarization (CMBP) constitutes
one of the major tools of the modern cosmology \cite{Melk1,Melk2};
its signal is supposed to be of the order of 10$\%$ or lower with
respect to the CMBR anisotropies \cite{Kosowsky,Hu-White,Melk3}.
In order to detect this faint signal, it is necessary to
characterize the instrumental spurious polarization. Once the
instrument is characterized, it is necessary to measure
calibration sources like planets and  HII regions, in order to
create a catalogue of polarized sources useful for all
polarization experiments \cite{melk4}.
\\
Moreover it is necessary to study the spectral polarization of the
foregrounds \cite{Hanany,foreground} measuring extensive sky
regions at different frequencies.
\\
During the last years some experiments have produced important
results about polarization of the CMBR; the WMAP satellite
experiment, in its first year data, has produced TE correlation
spectrum at 22, 33, 41, 61 and 94 GHz, and correlation maps for
small and large angular scales \cite{WMAP}. DASI, experiment
located at South Pole has produced the TE and EE correlation in a
range of frequencies between 26-36 GHz at multipoles 140-900
\cite{DASI_1,DASI_2}.
\\
The Polarimeter that we propose, MITOPol \cite{polarimetro},
intends to measure the polarization of the anisotropies of the
cosmic microwave background, the polarization induced on the CMBR
from S-Z effect and it aims to create polarized sourses maps in
the range of frequencies between 120-360 GHz with a 5 arcmin beam.
In this frequencies range, the polarized foregrounds contribution
is minimal and, at the same time, the CMB signal is maximum.

        \section{Experimental setup}

        MITOPol experiment is composed by three parts\cite{pisano1,pisano2}: a modulating system
         to discriminate
         the polarized signal from unpolarized part; a modified
         Martin-Puplett
        interferometer ($MPI$ here after) \cite{puplett,martin}, for spectral sampling
        in 4-12 $cm^{-1}$, and a cryostat with a $^{3}He$ cold stage where two bolometers are cooled
        down to a temperature
        of 0.3 K.
        \\
        MITOPol is a ground based experiment optically designed to
        be installed at the focal plane of MITO
        (\textit{Millimeter and Infrared Testagrigia Observatory }) telescope \cite{DePetris} situated
        on the Plateau Rosa (AO) at 3480 m a.s.l.
        \\
         The modulation of the polarized part of the signal can
        be realized by two different methods: an amplitude modulation
        with the Fresnel
        Double Rhomb ($FDR$ here after) or a phase modulation inside the $MPI$.
        Two different modulating methods have a
        different optical configuration:
        in the first optical configuration the image of the primary mirror is formed
        on the Winston cone aperture; therefore they perform an area selection of the observed
        portion of primary mirror, and an angle selection of
        observed sky.  In the second configuration the image of the sky is focused
        on the Winston cone aperture, so it selects in area the field of view
        and in angle the observed portion of secondary mirror.
        \\
        Both configurations ensure a troughput of 0.055 $cm^{2}$sr.

         \section{Martin-Puplett interferometer}

\begin{figure}
         \centering
        \includegraphics[width=0.6\textwidth,keepaspectratio]{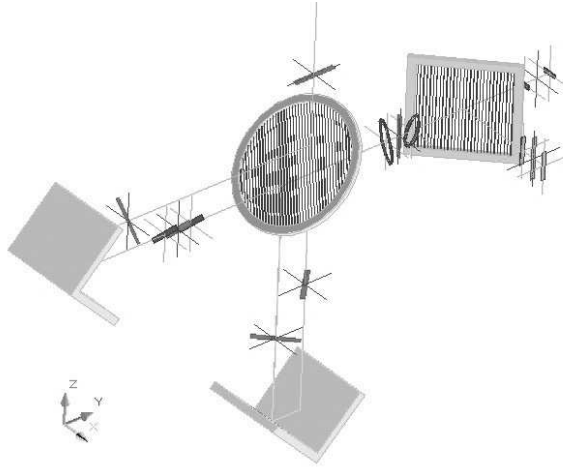}
        \caption{\textit{Schematic view of the Martin-Puplett interferometer.}}
        \label{Fig_1}
        \end{figure}

         The $MPI$ (Fig. \ref{Fig_1}) consists of a wire-grid, whose tungsten wires have a
         diameter of 10 $\mu m$ at the distance of 25 $\mu m$ with each other, and of
         two roof mirrors.
         The wire-grid is mounted so that the incoming radiation
         sees the wires under an angle of $45^{o}$ with respect to the interferometer optical axis.
         One of the two roof mirrors is fixed, while  the other is moved by a step motor;
         every step is 10 $\mu m$ and we can rich the maximum excursion of approximately 3.15 cm
         allowing as a spectral resolution of 0.16~$cm^{-1}$~\cite{Bell}.
         The incoming polarized radiation passes through the
         wire-grid,
         is splitted in its two orthogonal components and
         after the roof-mirror reflection is phase-shifted to $\pi /2$
         so that the transmitted radiation is now reflected and vice versa.
         \\
         We can write the ideal
         Martin-Puplett interferometer Mueller matrix as following:

          \begin{equation}
          \label{mmpi}
          M_{MP}= \left(
          \begin{array}{cccc}
            1 & 0 & 0 & 0 \\ 0 & \cos(\delta) & 0 & \sin(\delta) \\ 0 & 0 & -1
            & 0 \\ 0 & \sin(\delta) & 0 & -\cos(\delta)
            \end{array}\right)
            \end{equation}

            where $\delta$ represents the roof mirror phase-shift.

        \subsection{Cryostat}

\begin{figure}[h!]
        \centering
        \includegraphics[width=8cm]{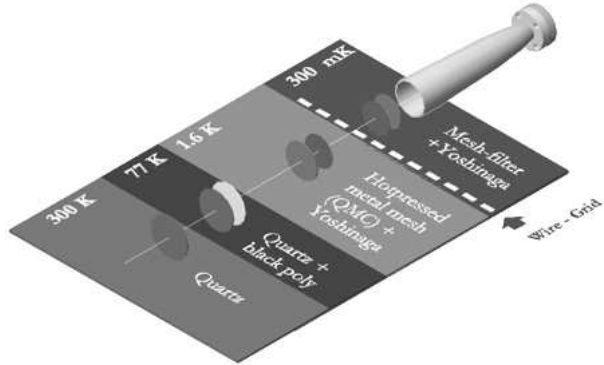}
        \caption{\textit{MITOPol filter chain.}}
        \label{Fig_2}
        \end{figure}

        The MITOPol cryostat
        is composed by two tanks, one for nitrogen (2.5 l) and one for the
        $^{4}$He (2.8 l), and a $^{3}$He fridge; the working temperature
        is approximately 300 mK with a duration of cooling cycle
        of approximately 18 hours. Inside the cryostat, in thermal
        contact with
        cryogenic stages, a filter chain is mounted as shown in Fig. \ref{Fig_2}.
        \begin{table} [h!]
        \label{filters}
        \begin{center}
        \begin{tabular}{|c|c|c|}
        \hline $\bf{Filters}$ & $\bf{Cut-off/Cut-on (cm^{-1})}$ & $\bf{Temperature (K)}$ \\
        \hline Quartz & Low-pass (100) & 300 \\
        \hline Quartz + black poly & Low-pass (400) & 77 \\
        \hline Yoshinaga & Low-pass (55) & 1,6 \\
        \hline Mesh & Low-pass (14) & 1,6 \\
        \hline Yoshinaga & Low-pass (50) & 0,3 \\
        \hline Mesh & Low-pass (12) & 0,3 \\
        \hline Winston cone & High-pass (4) & 0,3 \\
        \hline
        \end{tabular}
        \end{center}
        \caption{\textit{MITOPol filters chain with the relative working temperatures. The final band is
        4-12 $cm^{-1}$.}}
        \end{table}
A wire-grid splits the light in two beams collected by the f/3.5
Winston cones and absorved by two bolometers. The presence of two
channels, operating at the same frequencies, allows a more
efficient offset and noise removal. The incident signal on the
detectors can be represented in the following way:

\begin{equation}
I_{1,2}=I_{0}\pm(Q*f(w,\delta,t)+U*g(w,\delta,t))\mp V\sin\delta
\end{equation}

where $I_{0}$,$Q$,$U$,$V$ are the Stokes parameters, $\delta$ is
the roof mirror phase-shift and $f$,$g$ are two periodic functions
that depend from the chosen modulation system.\\ Observing the
difference of the two outputs, normalized for the sum, we can
detect the polarized signal embedded in the strong unpolarized
one. The bolometers used in this experiment are spider-web
bolometers developed at the University of Cardiff \cite{Angiola}.
We use differential elettronic read-out in order to reduce DC
components, microphonic noise and correlation between channels; we
use 2 JFET to common drain, mounted on the $^{4}$He stage heated
at a temperature of 120 K. Experimentally a value of $\sim
3nV/\sqrt{Hz}$ has been measured, optimal for the high sensibility
demanded. The incident background (in the best atmospheric
conditions) on the bolometers is of 500 pW, considering all the
transmission curves of the filters; therefore the bolometer
thermal conductivity has been chosen of the order of 10$^{-9}
W/K$.

\section{Modulation System}

The modulating element is fundamental in an experiment that aims
to measure the polarized part of a signal, in fact it can alter
the stage of the polarized signal, while leaving the unpolarized
unaltered allowing to separate the polarized light from the
unpolarized part of it.\\ In the following we will consider two
possible modulation systems.

\subsection{The Fresnel Double Rhomb}

   \begin{figure}[h!]
    \begin{center}
    \includegraphics[width=8cm]{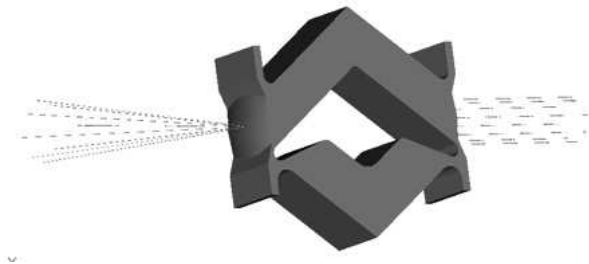}
    \end{center}
    \caption{\textit{Simulation of $FDR$}}
    \label{fig:Fig_3}
    \end{figure}

 $FDR$
 is an optical element obtained by single polyethylene HDPE (High Density Poly-Ethylene) block
 constituted by two rhomboidal base prisms with an angle of $106^{o}.6$ between them.
 This object is based on the internal total reflection.
 The internal walls of the $FDR$ are tilted to $53^{o}.3$ with respect to the optic
 axis of the system, therefore the incident light on them will be totally reflected
 since the critical angle of total internal reflection for the passage from polyethylene to air is $41^{o}.14$.
 The radiation inside the $FDR$ undergoes four
 reflection on its
 walls.
 Using the Fresnel's equations, it is possible to calculate, in case of total internal reflection,
 the phase-shift induced by a single reflection
 in this dielectric\footnote{The equations are obtained assuming that the optical
properties of two medium are determined
 from real refractive indices and that the materials are homogenous as in the separation surface
 as to the inside, therefore to avoid losses due to scattering.}:

 \begin{equation}
    \delta = \phi_s - \phi_p = 2 \arctan \frac{\cos \vartheta_1 \sqrt{\sin^2 \vartheta_1
    -(1/n)^2}}{\sin^2 \vartheta_1}
    \end{equation}

 where $n$ is the HDPE refraction index.
 \\
 From this equation, in order to obtain a
 phase-shift
 of $\pi/4$ we obtain two angles:
 \begin{equation}
  \vartheta_1=47^{o}.58
  \end{equation}
  \begin{equation}
 \vartheta_1^{' }=55^{o}.37
 \end{equation}

 The $FDR$ input (spherical shape with a curvature radius of 69 mm) is placed in the MITO f/4 focal
 plane \cite{DePetris} (the output has the same shape for simmetry)
 so inside the $FDR$ the beam is f/8.8. Using ray-tracing simulations we have
 obtained a distribution of phases-shifts centered around the value $\pi/4$ for an
 angle of $53^{o}.3$.
  \\
  To modulate the polarization, we rotate the $FDR$
  at a set frequency ($\omega$).
  Using Stokes formalism, the ideal $FDR$ Mueller Matrix becomes:

  \begin{equation}
    \mbox{{\bfseries M}}_{DR} =
    \left(
    \begin{array}{cccc}
    1               &       0        &       0         & 0 \\
    0               & \cos4\omega t  & \sin4\omega t   & 0 \\
    0               & \sin4\omega t  & -\cos 4\omega t  & 0 \\
    0               & 0              & 0               & -1
    \end{array}
    \right)
    \end{equation}

   Therefore $Q$ parameter is modulated with a frequency 4 times the mechanical one, since bolometers are only sensitive
   to the intensity of the light, we add a
   wire-grid at the output of the $FDR$ to analyze the polarization of radiation; on the contrary the unpolarized radiation
   passes through unchanged.
   \\
   The advantages of the $FDR$
   with respect to the other modulation techniques are multiple: first of all the total reflection
   is not dispersive,
   so has a wide spectrum application;
   the total internal reflection does not attenuate the signal but it produces a phase-shift in the 4
   reflections. Finally the polarization is modulated at a double frequency
   compared to the mechanic one allowing an efficient removal of microphonic noise connected to
   the measures.

   \subsubsection{Efficiency tests}

   \begin{figure}[b!]
    \centering
    \includegraphics[width=6.5cm, keepaspectratio]{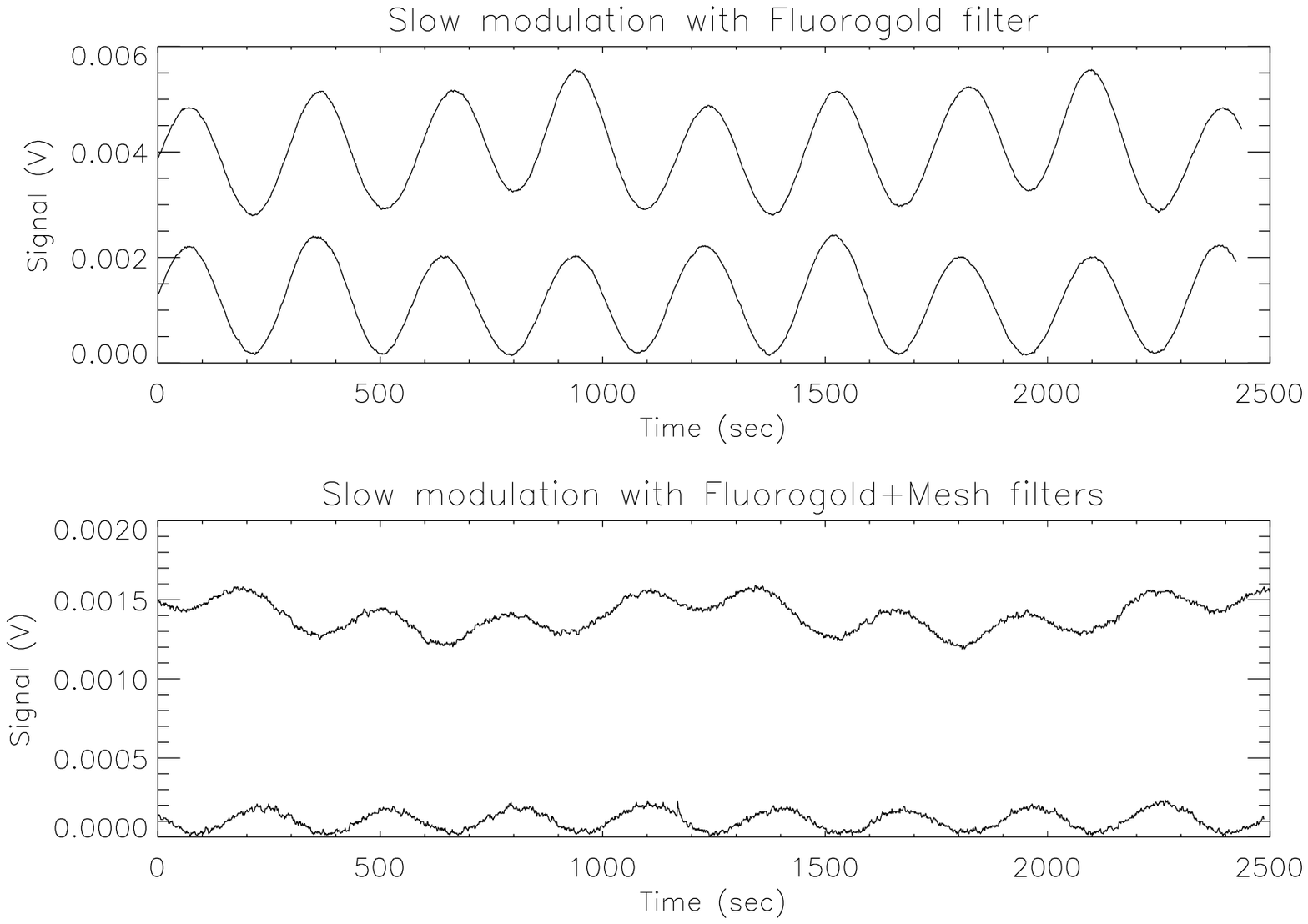}
    \includegraphics[width=6.5cm, keepaspectratio]{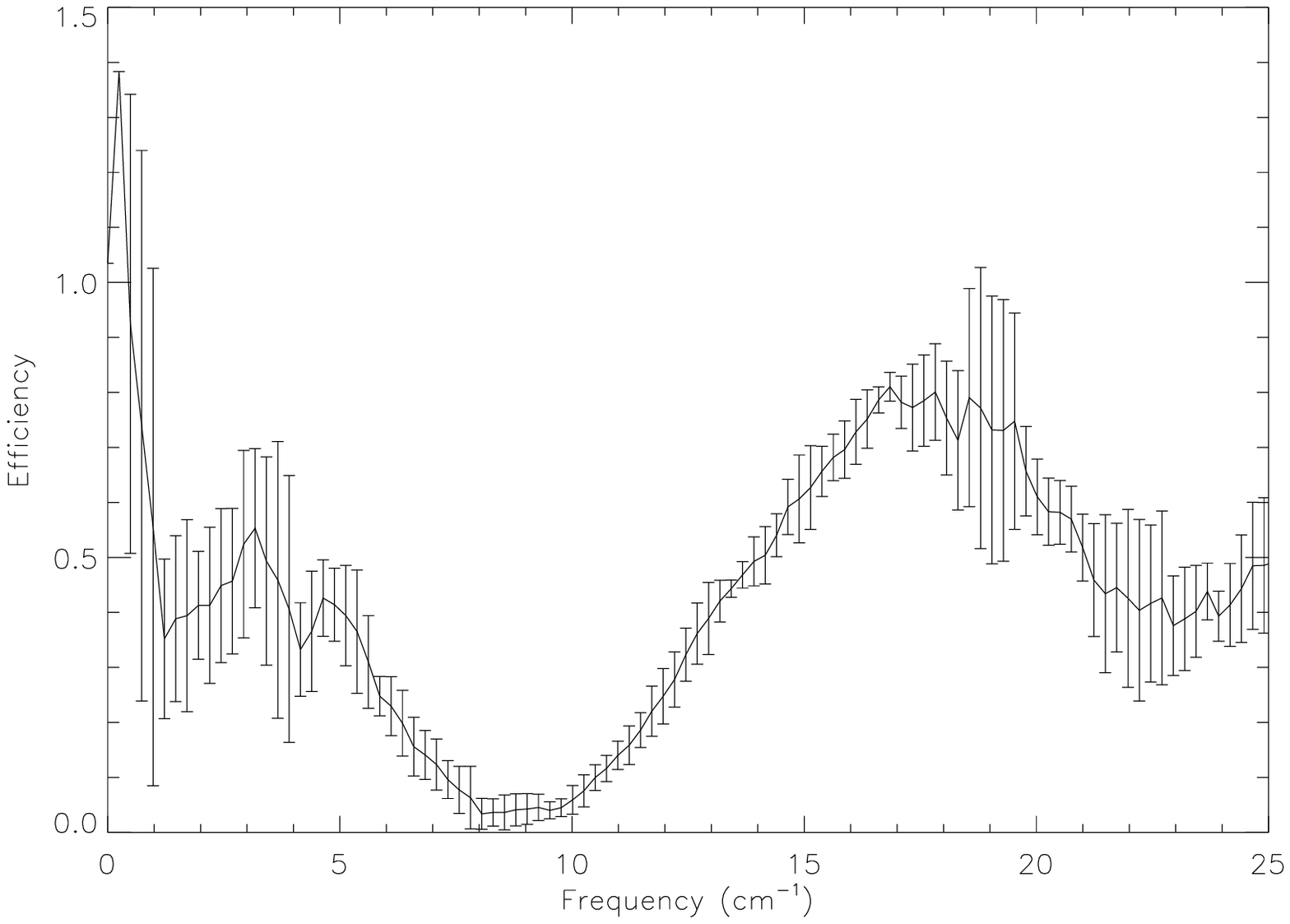} \\
    \caption{\textit{Slow modulation data from measurements with two different setup: in the first
    (top-left panel) a fluorogold filter
    (with a cut-off about 30 $cm^{-1}$) is used. In the second (bottom-left) we reduced the band by means
    of a low-pass Mesh filter with cut-off to
    10 $cm^{-1}$ in series with the fluorogold filter. In each of these plots, two curves are presented, one with
    parallel wire-grids and a second one with wire-grids orthogonal with each other.
    The right diagram represents the ratio of the two spectra
    measured at two maximum positions in the slow modulation configuration with
    parallel and orthogonal wire-grids using only the fluorogold filter. The efficiency in our band is $19\pm5$\%.}}
    \label{fig:Fig_4}
    \end{figure}

   Efficiency tests have been performed illuminating the $FDR$ with a polarized radiation and detecting it as it
   gets modified by th $FDR$ itself.
   The tests are realized by using a Hg lamp as a source.  The lamp emits in all the electromagnetic
   spectral range and in the millimetric its emission is similar to a black
    body at a temperature between 3000 K and 3500 K. Moreover a Lamellar Grating
    interferometer \cite{Bell}
    has been used to measure the spectral behaviour of the $FDR$.
    \\
    Several interferograms have been realized with 256 mechanical steps of
    80 $ \mu m$ each one. The frequency band-width can be investigated
    in a range between 3-31.25 $cm^{-1}$.
    \\
    A lab. cryostat has been used for laboratory tests
    using a bolometer working at 1.6 K temperature.
    \\
    We set the outgoing beam from the
    Lamellar Grating as an f/4, since it must reproduce
    as faithfully as possible the working conditions of the instrument.

    In order to study its polarization efficiency,
    we have used two wire-grids: the first one located in front of
    $FDR$
    and the second one behind it.
    In this situation the
    radiation that enters into the $FDR$ is completely polarized in the orthogonal direction respect to the
    wires of the wire-grid; into the $FDR$ the
    radiation is phase-shifted, and then the second wire-grid
    allows to analyze the modulated contribution of polarization.
    If the $FDR$ rotates at a frequency about 2 mHz, a modulation is observed at a frequency 4
    times the mechanical one. Ideally, if we perform two different measurements one with parallel wire-grids
    and the second one with orthogonal wire-grids with each other, we should observe
    the same amplitude in both signals with a phase-shift of $\pi/2$.
    Then we made two interferograms stopping the $FDR$ in two positions corresponding to the maximum
    of modulation in the parallel and orthogonal wire-grids
    setups.
    The expected result is a constant spectrum with unity value.
    Fig. \ref{fig:Fig_5} evidences a loss of efficiency in the centre of our band-width; this corresponds to an
    efficiency of phase-difference in the band 4-12 $cm^{-1}$ of $19\pm5$\%.
    \\
    To explain this inefficiencies different motivations have been
    proposed and investigated:
    diffractional effects could change the beam f/\#. Gaussian Optic
    simulation have been shown that inside the $FDR$ the effect is
    negligible. In any case we would expect a global decrease of
    the efficiency at low frequencies which is not observed in
    our data.
     \\
    The possibility for a radiation beam to go through the
    structure housing the $FDR$ without entering the
    $FDR$ itself has been ruled out by several
    measuraments without the external structure and shielding
    with aluminium and Eccosorb all the possible leaks.
    \\
    Tests have also been performed using a more collimated beam
    with respect to the f/\# that the $FDR$ would see at
    the telescopes focal plane showing the same results as in
    previous case.
    \\
    The possibility that the polyethylene (the one we have used) has
    a varying refraction index with the frequency has been
    investigated by measuring it in the already cited frequency
    bands.
    We have used, as source,
    Eccosorb at 77~K and 300 K modulated at a frequency of 12 Hz and as interferometer the $MPI$.
    \\
    This measurement has been performed by placing a polyethylene
    sample in one of the arms of
    $MPI$
    and measuring the distance between the zero path difference position obtained with and without
    the sample itself. This has allowed to measure the optical
    path delay introduced by polyethylene and, known the sample
    thickness, one can derive the refraction index.
    \\
     The difference is equal to:

     \begin{equation}
     \Delta^{ZPD}_{Opt}=d(n-1)
     \end{equation}

     Where $\Delta^{ZPD}_{Opt}$ is the optical shift, $d$ is the polyethylene sample thickness
     and $n$ is the polyethylene refraction index.
     \\
     This measurement realized using band-pass filters inside our spectral range, has
     confirmed refraction index value reported in lecterature. The integrated
     value inside the band is:
    \begin{equation}
     n=1.5276\pm0.0066
    \end{equation}
       \begin{figure}[h!]
      \begin{center}
      \includegraphics[width=7cm]{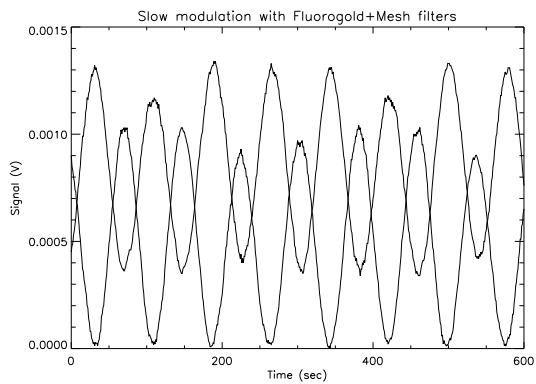}
      \end{center}
      \caption{\textit{Slow modulation with the same setup of Fig. \ref{fig:Fig_4} using the
      Mesh filter with cut at 10 $cm^{-1}$ and the fluorogold filter in series: the improvement
      is evident.}}
      \label{fig:Fig_6}
      \end{figure}
    \begin{figure}
      \begin{center}
      \includegraphics[width=7cm]{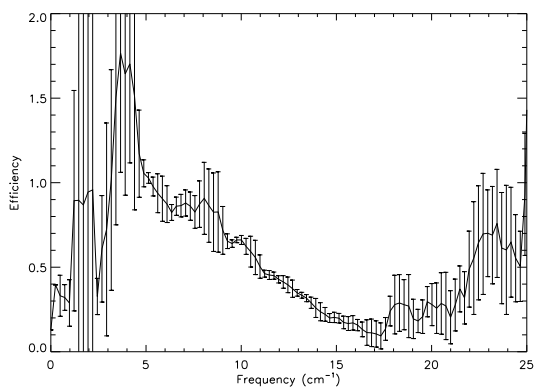}
      \end{center}
      \caption{\textit{Spectrum obtained with the same setup of Fig. \ref{fig:Fig_4}:
      the loss of efficiency towards greater
      frequencies. In this case the phase-shift efficiency is increased to $86\pm8$\%.} }
      \label{fig:Fig_7}
      \end{figure}
A further possibility is that our $FDR$ could be affected by
optical activity since, at manufacturing stage, it has undergone
stresses and thermal shocks \cite{ALMA_1,ALMA_2}. The optical
activity produces, in analogy to the birifrangent materials, an
induced polarization \cite{Cantor}. It depends by the molecule
simmetry that constitutes the substance and by the degree of
disorder that is inside the lattice. All the molecules of organic
nature or synthetize by living organism, are optical active. The
structure of the polyethylene is constituted with a carbon chain
where every atom is tied to two hydrogen atoms; if the chain is
linear it is defined high density polyethylene (HDPE). These
chains can be very long assuming macroscopic dimension and
therefore comparable with the wavelengths that we investigate. As
a result, we could see anisotropies and spurious polarization
effects varying with the frequency. This effect has been tested by
heating the $FDR$ at temperatures just below the HDPE melting
point (137 $C$) in order to let the internal structure relax and
acquire again the isotropic structure needed for an optical
element to be used in polarization measurements. The $FDR$ has
been gradually heated and finally left at 135 $C$ for 48 hours and
at 137 $C$ for 12 hours.
\\
After these thermal cycles we have repeated the same measurements
in order to compare the results which are shown in Fig.
\ref{fig:Fig_6} and Fig.~\ref{fig:Fig_7}.
\\
The $FDR$ efficiency is considerably increased ($86\pm8$\%).

\subsection{Phase modulation with Martin-Puplett interferometer}

Using $MPI$ is an alternative method to modulate the polarization,
both as an interferometer and as a phase modulator
\cite{chamberlain,chamberlain2}. The basic idea is to wobble one
of the two roof mirrors along the optical axis. In order to write
the $MPI$ Mueller matrix one needs to consider eq. \ref{mmpi} and
to substitute $\delta\rightarrow\delta^{'}=\delta - \delta_{m}$,
where $\delta= 2 \pi \Delta x_{opt}/ \lambda$ and $\delta_{m}= 2
\pi f(\omega,t)/\lambda$.
\\
The optical path difference $\delta^{'}$ represents the modulation
term. The function f($\omega$,t) represents the roof mirror wave
form. The new Martin-Puplett Mueller matrix becomes:

\begin{equation}
\label{nuovammp}
M_{MP}^{New}= \left(
\begin{array}{cccc}
1 & 0 & 0 & 0 \\ 0 & \cos(\delta^{'}) & 0 & \sin(\delta^{'}) \\ 0
& 0 & -1 & 0 \\ 0 & \sin(\delta^{'}) & 0 & -\cos(\delta^{'})
\end{array}\right)
\end{equation}

From eq. \ref{nuovammp} we can note that the polarizated radiation
is modulated while the unpolarized radiation remains unchanged;
however the polarization plane is not rotated. The effect of the
modulation is independent from the choice of the oscillating roof
mirror; the practical solution that we adopted has been of
oscillating  the roof mirror each step.
\\
Using a lock-in amplifier, the output is proportional to the
interferogram derivative. The smaller is the amplitude the more
the signal approximates to punctual derivative but with a
decreasing intensity. Nevertheless one needs to consider that the
radiation wavelength is in the range between 850 $\mu m$ and 2.5
$mm$ so, in order to be efficiently modulated, it is not possible
to choose an amplitude modulation much smaller than the wavelength
of interest. On the other side, important spectral information can
be lost if we modulate with an amplitude higher than the step; as
a matter of fact using a great amplitude could be possible to
modulate among two points with similar intensity, smoothing
therefore the interferogram. An optimal choice for the amplitude
modulation is to set it equal to the modulation step.

\begin{figure}[!htbp]
\centering
\includegraphics[width=0.6\textwidth,keepaspectratio]{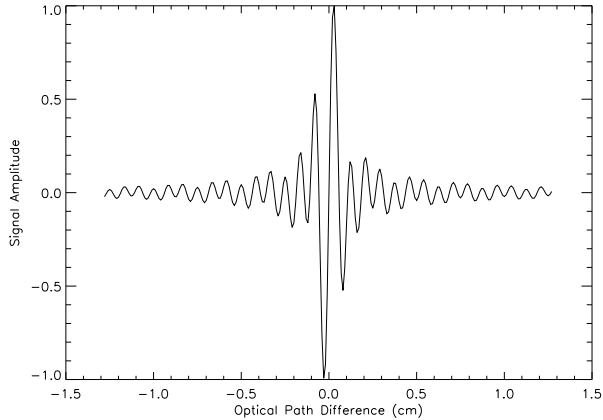}
\caption{\textit{Simulated signal obtained with phase modulation
.}}\label{confrontoderiv1}
\end{figure}

The relation that links Fourier transform of interferogram to its
derivative is:

\begin{equation}
\label{foubes} FT[f^{'}(x)](k)=C_{o}\cdot |2J_{1}(2\pi kA/2)|\cdot
FT[f(x)](k)
\end{equation}

where in the case of phase-modulation $C_{o}$ is the term of
Fourier series, \textit{A} is the modulation amplitude and $J_1$
is the Bessel first kind and order 1 function.

From eq. \ref{foubes} we note that the result is always lower than
1, compared with ``classical'' modulation; then, to maximize the
signal, it is worth to choose the wave form in order to have the
maximum $C_{o}$. The wave square modulation is the best with
$C_{square}=4/\pi$.
\\
One of the classical spectroscopy problems, particularly at high
resolution, is the difficult to recognize between small variations
due to the source and variations due to other factors; in fact a
small variation on the interferogram, due for example to
atmospheric fluctuations, affects heavily the spectrum, and
produces signal variations that may be confused with emission or
absorption lines \cite{Maillard}.
\\
Using the phase modulation this problem has been solved; in this
case, the baseline of the interferogram is about zero, see Fig.
\ref{confrontoderiv1}; instead the baseline of the classical
modulation is $I(0)/2$ \cite{Mertz}; and so any signal originated
by a fluctuation (instantaneous) is near to the zero level of the
interferogram and does not affect the spectrum.
\\
Another advantage of phase modulation, is observation time; in
fact with phase modulation we observe constantly the source.
Moreover \cite{chamberlain,chamberlain3}, the signal to noise
ratio is $\frac{(S/N)_{PM}}{(S/N)_{AM}}=4\mid J_{1}(2 \pi \nu
\frac{A}{2})\mid$, and so when the Bessel function is greater than
0.25 the signal to noise ratio, in the phase modulation (PM), is
greater than the amplitude modulation (AM) \cite{ade}. The
spectrum obtained with the phase modulation is divided by Bessel
function (first kind and order 1) thus, when this function is
equal to zero, we have a loss of informations. While the amplitude
modulation is small there are no problems; when the amplitude
modulation is high, this zero value could be on the frequency
range that we are studying. The loss of information, choosing
amplitude modulation equal to the interferogram step, is out of
our range.

\subsubsection{Measurement}

\begin{figure}[h!]
\centering
\includegraphics[width=13cm,keepaspectratio]{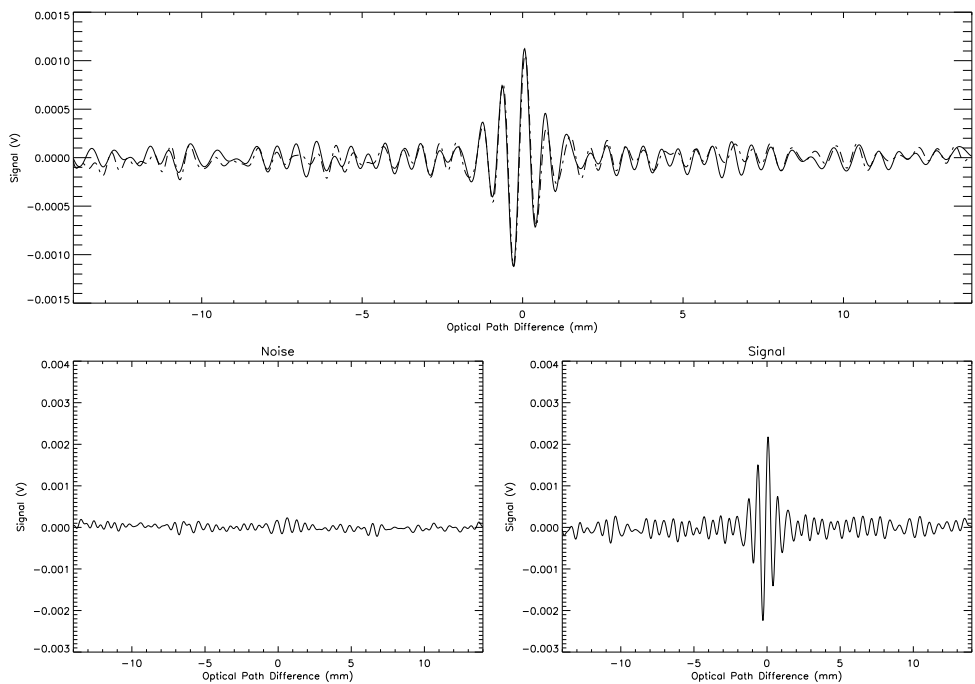}
\caption{\textit{Interferogram obtained at $60^{o}$ of elevation
(the two signals are overlapped); we note the good correlation
between the two channels.}}\label{mega}
\end{figure}

\begin{figure}[h!]
   \centering
   \includegraphics[width=6.5cm,keepaspectratio]{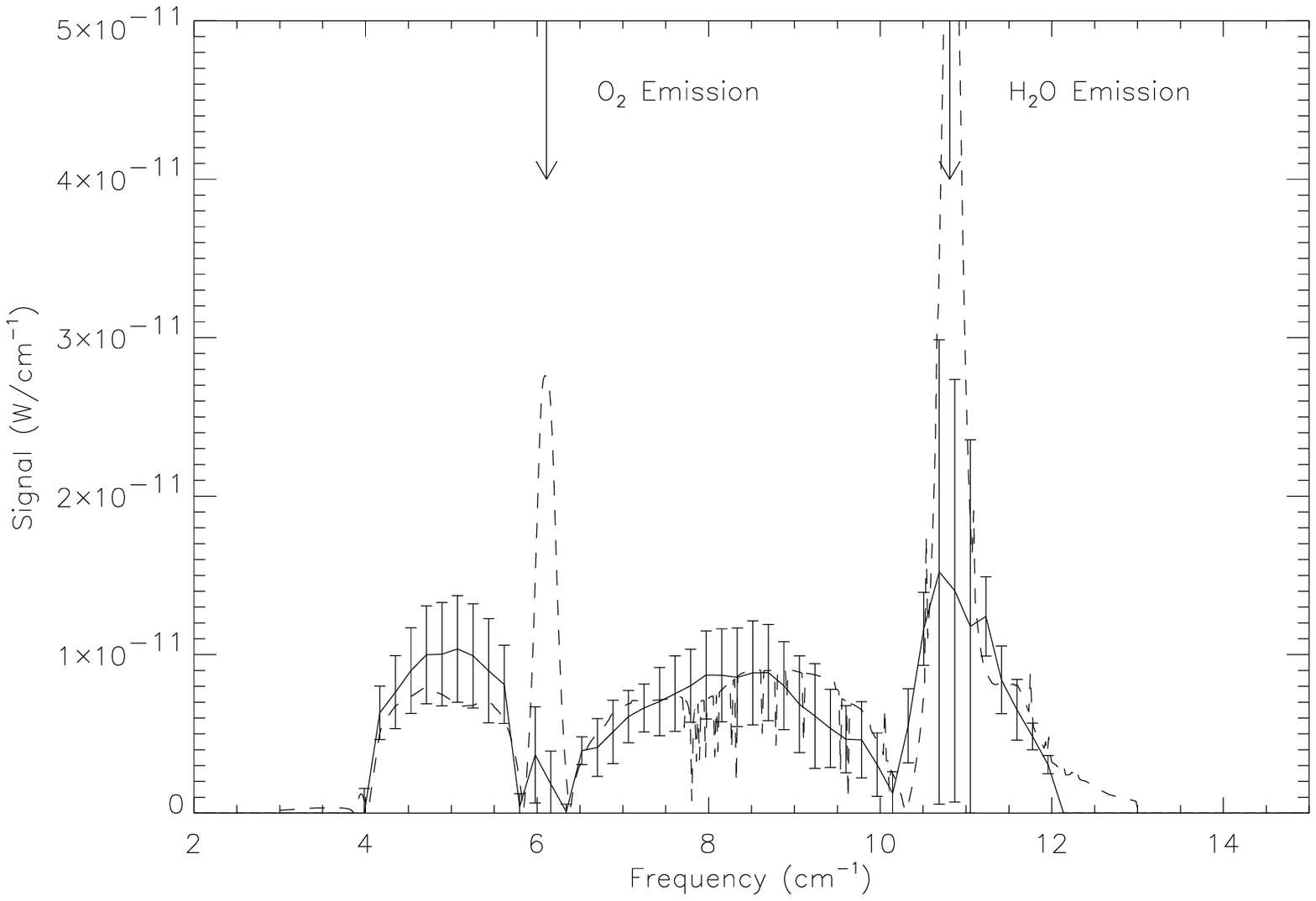}
   \includegraphics[width=6.5cm,keepaspectratio]{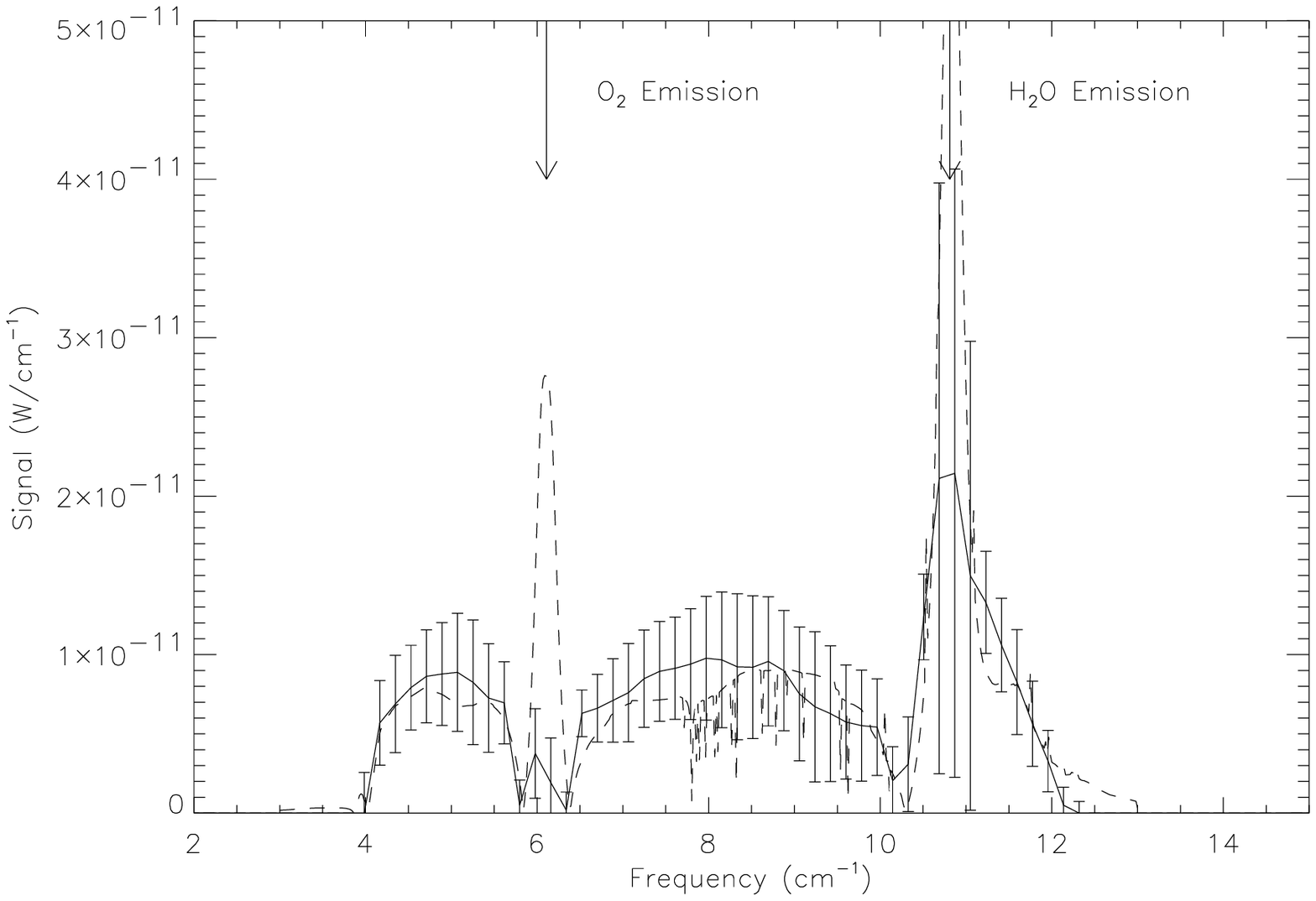} \\
   \caption{\textit{Atmospheric emission at MITO (subracted by a Black Body source at 77K as reference source),
   measured with two channel of the polarimeter. The outlined line is an atmospheric emission
   simulated with pwc 2.3 mm. It is clear the loss of spectral resolution at frequencies corresponding to $H_2O$ and $O_2$ emission}}
   \label{atm12}
   \end{figure}

\begin{figure}[h!]
\centering
\includegraphics[width=0.8\textwidth,keepaspectratio]{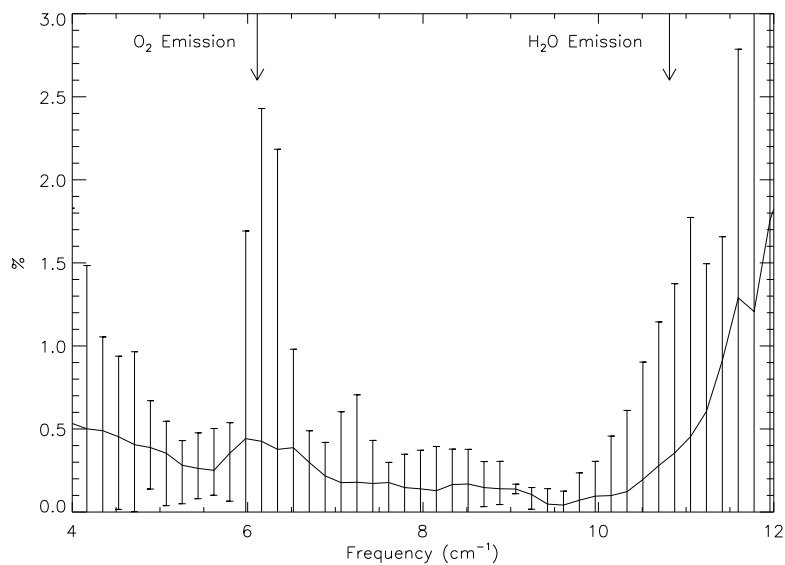}
\caption{\textit{Instrumental spurious polarization in one
direction (Telescope + Polarimeter).}} \label{Polsp}
\end{figure}

We have mounted this experiment at focal plane of MITO telescope
\cite{DePetris2}, we have obtained spectra on atmospherics
emission in the range 4-12 $cm^{-1}$, placing the wire-grid in
front of the interferometer polarizing all the incident radiation.
\\
In Fig. \ref{mega} we have reported the interferograms operating
on the atmosphere emission, obtained with an elevation equal to
$60^{o}$ with step and amplitude modulation equal to 100 $\mu m$.
From these interferograms we have obtained the spectrum, shown in
Fig. \ref{atm12} and we have compared it with simulated emission
with ATM program  \cite{Guelin,pardo,cerni,pardotesi}.
We have performed polarimetric measurements removing the wire-grid
in front of Martin-Puplett interferometer; any excess of
polarization obtained from this interferograms is an indicator of
the spurious polarization of the instrument, see Fig. \ref{Polsp},
considering that the atmosphere emission at this wavelength is not
expected to be polarized \cite{Hanany}.
From the Fig. \ref{Polsp} we obtain a value of the spurious
polarization in one direction lower than 1\%.

\section{Conclusions}
In this paper we have presented two polarization modulation
methods. The first method have evidenced that the optical activity
of the polyethylene realized by a not homogeneous block can alter
the entering polarization; on the other side this amplitude
modulation method is independent by the wavelength of the
radiation.
\\
The phase modulation gains observation time and does not present a
decrease by transmission. However the polarization plane is not
rotate so we have to insert a optic element that produces a
polarization rotation after the interferometer.

\section*{Acknowledgments}
The authors thank Dr. Giampaolo Pisano for his extensive work in
this project during the past years.

\newpage

\begin{flushleft}
{\Large {\bf Appendix}}
\end{flushleft}

\section*{An introduction to the Stokes parameters}

    A plane electromagnetic wave with its Poynting vector directed along the \textit{z} axis can be
    decomposed into its two components in the \textit{x} and \textit{y} direction: $E_{x}(z, t)$ and $E_{y}(z,t)$.
   If any correl
   ation between two vectors exists, the plane wave will be defined
    polarized.
 \\
 In order to obtain observable quantities,
 we always have to consider the temporal average of the polarization ellipse.
 However the temporal process of average can be avoided
 representing
 the real optical amplitudes in terms of complex amplitudes:

    \begin{eqnarray}
        I &=& E_x E_x ^* + E_y  E_y ^* \nonumber \\
        Q &=& E_x E_x ^* - E_y  E_y ^* \nonumber \\
        U &=& E_x  E_y ^* + E_y  E_x ^*  \\ \label{eqn:stokes}
        V &=& E_x E_y ^* - E_y  E_x ^* \nonumber
        \end{eqnarray}

    These quantities are called {\bfseries Stokes parameters};
    \textit{Q} and \textit{U} depend, for construction,
    on the chosen coordinates system.
    \\
    \textit{I}  represents the intensity of the e-m wave considered.
    \\
    \textit{Q}  represents the contribution of the linear polarization horizontal and
    vertical.
    \\
    \textit{U}  describes the contribution of linear polarization at an angle of
    $\pm45°$.
    \\
    \textit{V}  describes the contribution of circular polarization towards right and
    left.
    \\
    The Stokes parameters are real quantities, they are observable
    and the following expression is always true:

     \begin{equation}
     I^2 \geq Q ^2 + U ^2 + V ^2
     \end{equation}

     where the equal is used for fully polarized light.
     It is possible to define a Stokes vector
     whose 4 parameters are the elements of a column vector.

      \begin{equation}
        {\mathcal S} = \left(
        \begin{array}{ccc}
        I \\
        Q \\
        U \\
        V
        \end{array}
        \right)
      \end{equation}

     The Stokes parameters describe the wave polarization degree;
     this can be evidenced defining the quantity:

      \begin{equation}
      P \equiv \frac{I_{pol}}{I } = \frac{\sqrt{Q ^2 + U ^2 + V ^2}}{I },\  0 \leq P \leq 1
      \end{equation}

        When a beam crosses an element that can change its polarization state,
        the elements of the Stokes vector will vary depending on the particular
        considered element.
        \\
        We consider an incoming beam defined by whichever Stokes vector;
        the outgoing beam from the element will be characterized by a vector
        $S_i^{'}=(I^{'},Q^{'},U^{'},V^{'})$ where

        \begin{eqnarray}
        I ^{'} &=& m_{00} I + m_{01} Q + m_{02} U + m_{03} V \nonumber \\
        Q ^{'} &=& m_{10} I + m_{11} Q + m_{12} U + m_{13} V \nonumber \\
        U ^{'} &=& m_{20} I + m_{21} Q + m_{22} U + m_{23} V  \\
        V ^{'} &=& m_{30} I + m_{31} Q + m_{32} U + m_{33} V \nonumber
        \end{eqnarray}

        therefore the following equation can be written:

        \begin{equation}
        S^{'} = M \cdot S
        \end{equation}

        M is the Mueller matrix of the optical element.
        The electric field vector of a e-m wave can be changed in its amplitude,phase or direction
        and this change is described by the matrix M.
        \\
        The terms in a Mueller matrix have different meanings:
        the terms
        found in the trace define the transmission and the depolarization of the element.
        The terms $m_{01}, m_{02}, m_{03}, m_{10}, m_{20}, m_{30}$
        contain the contribution
        of spurius polarization of the element.  The remaining terms contain the passage of the
        light from \textit{Q} to \textit{U} or \textit{V} polarization.
        \\
        On the contrary
        the inverse is not valid, as it is not possible to identify
        a coefficient by a single effect \cite{Collett,Huard}.



\begin{thebibliography}{}

\bibitem{Melk1}
C.Ceccarelli et~al.
\newblock {The Polarization of the Cosmic Background an the universal magnetic
  field}.
\newblock {\em Proceedings of the Seventeenth Moriond Astrophysics Meeting},
  A84:191--203, 1982.

\bibitem{Melk2}
N.Caderni et~al.
\newblock {Polarization of the Microwave Background Radiation I-Anisotropic
  cosmological expansion and evolution of the polarization states}.
\newblock {\em Physical Rev.D}, 17:1901--1907, 1978.

\bibitem{Kosowsky}
A.Kosowsky.
\newblock {Introduction to Cosmic Microwave Background}.
\newblock {\em New Astron.Rev.}, 43:157, 1999.

\bibitem{Hu-White}
W.Hu e~M.White.
\newblock {A CMB Polarization Primer}.
\newblock {\em New Astron.}, 2:323, 1997.

\bibitem{Melk3}
N.Caderni et~al.
\newblock {Polarization of the Microwave Background radiation II-an infrared
  survey of the sky}.
\newblock {\em Physical Rev.D}, 17:1908--1918, 1978.

\bibitem{melk4}
A.Blanco et~al.
\newblock {Polarization of the Cosmic Background Radiation - an experimental
  approach}.
\newblock {\em Marcel Grossmann Meeting: General Relativity}, 2D:919, 1982.

\bibitem{Hanany}
S.Hanany P.Rosenkranz.
\newblock {Polarization of the atmosphere as a Foreground for Cosmic Background
  Polarization experiments}.
\newblock {\em New Astron.Rev}, 2003.

\bibitem{foreground}
Angelica de~Olivera-Costa~et al.
\newblock {The Large-Scale Polarization of the Microwave Foreground}.
\newblock {\em Phys.Rev}, D:68, 2003.

\bibitem{WMAP}
A.Kogut et~al.
\newblock {Wilkinson Microwave Anisotropy Probe (WMAP) First Year Observations:
  TE Polarization}.
\newblock {\em ApJ.Suppl.}, 148:161, 2003.

\bibitem{DASI_1}
J.Kovac et~al.
\newblock {Detection of Polarization in the Cosmic Microwave Background Using
  DASI}.
\newblock {\em Nature}, 420:772--787, 2002.

\bibitem{DASI_2}
E.M.Leitch.
\newblock {Measuring Polarization with DASI}.
\newblock {\em Nature}, 420:763--771, 2002.

\bibitem{polarimetro}
E.S.Battistelli et~al.
\newblock {Far Infrared Polarimeter with Very Low Instrumetal Polarization}.
\newblock {\em SPIE conference proceedings:Astronomical telescopes and
  instrumentation}, 4843,241-249, 2003.

\bibitem{pisano1}
G.Pisano.
\newblock {Millimiter CBR polarimetry: the POLCBR experiment at MITO}.
\newblock {\em New Astr.Rev}, 43:329--339, 1999.

\bibitem{pisano2}
G.Pisano.
\newblock {Realizzazione e Calibrazione di un Polarimetro per misure della
  Radiazione di Fondo Cosmico nel lontano infrarosso}.
\newblock {\em PHD Thesis in Astronomy XII course, University of Rome 'La
  Sapienza'}, 2004.

\bibitem{puplett}
D.K.Lambert P.L.Richards.
\newblock {Martin-Puplett interferometer: an analysys}.
\newblock {\em Infrared Physics}, 17:1595, 1978.

\bibitem{martin}
D.H.Martin E.Puplett.
\newblock {Polarised interferometric spectrometry for the millimetre and
  submillimetre spectrum}.
\newblock {\em Appl.Opt.}, 10:105, 1969.

\bibitem{DePetris}
M.~De Petris et~al.
\newblock {A ground based experiment for CMBR anisotropy observations: MITO}.
\newblock {\em New Astron.Rev.}, 43:297--315, 1999.

\bibitem{Bell}
R.J.Bell.
\newblock {\em Introductory Fourier Trasform Spectroscopy}.
\newblock Academic Press, 1972.

\bibitem{Angiola}
A.Orlando.
\newblock {Optimization and realization of bolometric detectors for high
  sensitivity measurements of the Cosmic Microwave Background}.
\newblock {\em PHD Thesis in Astronomy XVI course, University of Rome 'La
  Sapienza'}, 2004.

\bibitem{ALMA_1}
G.A.Ediss and D.Koller.
\newblock {68.5 to 118 GHz measurament of possible infrared filter materials:
  black polyethylene,zitex and grooved and un-grooved fluorogold and HDPE}.
\newblock {\em ALMA MEMO}, N412, 2002.

\bibitem{ALMA_2}
G.A.Ediss and T.Globus.
\newblock {60 to 450 GHz trasmission and reflection measuraments of gooved and
  un-grooved HDPE plates}.
\newblock {\em ALMA MEMO}, N347, 2001.

\bibitem{Cantor}
C.Cantor and P.Schimmel.
\newblock {\em Biophysical Chemistry}.
\newblock W.H.Freeman, 1980.

\bibitem{chamberlain}
J.Chamberlain.
\newblock {Phase modulation in far infrared (submillimetre-wave)
  interferometers. I-Mathematical Formulation}.
\newblock {\em Infrared Physics}, 11:25--55, 1971.

\bibitem{chamberlain2}
J.Chamberlain H.A.Gebbie.
\newblock {Phase modulation in far infrared (submillimetre-wave)
  interferometers. II-Fourier Spectrometry and Terametrology}.
\newblock {\em Infrared Physics}, 11:57--73, 1971.

\bibitem{Maillard}
P.Connes J.P.Maillard and J.Connes.
\newblock {Spectroscopie astronomique par transformation de Fourier}.
\newblock {\em Journal de Physique}, Colloque C2, Tome 28, pp.120-135, 1967.

\bibitem{Mertz}
L.Mertz.
\newblock {Spectromètre stellaire multicanal}.
\newblock {\em Journal de Physique et le radium}, Tome 19, pp.233-236, 1958.

\bibitem{chamberlain3}
J.Chamberlain M.J.Hine, J.Haigh.
\newblock {Phase modulation in far infrared (submillimetre-wave)
  interferometers. III-Laser Refractometry}.
\newblock {\em Infrared Physics}, 11:75--84, 1971.

\bibitem{ade}
J.E. Harries and P.A.R. Ade.
\newblock {The high resolution millimetre wavelength spectrum of the
  atmosphere}.
\newblock {\em Infrared Physics}, 12:81--94, 1972.

\bibitem{DePetris2}
M.~De Petris et~al.
\newblock {MITO: the 2.6 m millimeter telescope at Testa Grigia}.
\newblock {\em New Astron.Rev.}, 1:121, 1996.

\bibitem{Guelin}
Michel Guelin.
\newblock {Atmospheric Absorption}.
\newblock {\em Proceedings of the workshop on the IRAM Millimeter
  Interferometry Summer School}, 1998.

\bibitem{pardo}
E.Eerabin et~al.
\newblock {Submillimeter FTS Measurements of Atmospheric Opacity above Mauna
  Kea}.
\newblock {\em Appl.Opt.}, 37:2185--2198, 1998.

\bibitem{cerni}
J.~R.~Pardo E.~Serabyn, J.~Cernicharo.
\newblock {Atmospheric Transmission at Microwaves (ATM): An Improved Model for
  mm/submm applications}.
\newblock {\em IEEE Trans. on Antennas and Propagation}, 49/12, 1683-1694,
  2001.

\bibitem{pardotesi}
J.R.Pardo-Carrion.
\newblock {Etudes de l'atmosph\`{e}re terrestre au moyen d'observations dans
  les longueurs d'onde millimetriques et submillimetriques}.
\newblock {\em Doctorat Europeen Universite Paris VI - Universidad Complutense
  de Madrid}, 1996.

\bibitem{Collett}
E.Collett.
\newblock {\em Polarized Light}.
\newblock Marcel, 1993.

\bibitem{Huard}
S.Huard.
\newblock {\em Polarization of Light}.
\newblock J.Wiley and Sons, 1997.

\end{thebibliography}
\end{document}